# The geometry of high-dimensional phase diagrams: I. Generalized Gibbs' Phase Rule

Wenhao Sun[1*], Matthew J. Powell-Palm,[2,3] Jiadong Chen[1]
[1]Department of Materials Science and Engineering, University of Michigan, Ann Arbor, MI, 48109, USA
[2]J. Mike Walker '66 Department of Mechanical Engineering, Texas A&M University, College Station, TX, 77802, USA
[3]Department of Materials Science & Engineering, Texas A&M University, College Station, TX, 77802, USA

*Correspondence to: whsun@umich.edu

## Significance Statement

Phase diagrams are essential tools of the materials scientist, showing which phases are at equilibrium under a set of applied thermodynamic conditions. Essentially all phase diagrams today are two dimensional, typically constructed with axes of temperature-pressure or temperature-composition. For many modern materials, it would be valuable to construct phase diagrams that include additional forms of thermodynamic work—such as elastic, surface, electromagnetic or electrochemical work, *etc.*—which grows the free energy of materials into higher (≥3) dimensions. Here, we extend Gibbs' original arguments on phase coexistence to derive a generalized Phase Rule, based in the combinatorial geometry of high-dimensional convex polytopes. The generalized Phase Rule offers a conceptual and geometric foundation to describe phase boundaries on high-dimensional phase diagrams, which are relevant for understanding the stability of modern materials in complex chemical environments.

## Abstract

We revisit Gibbs' arguments on the equilibrium of heterogeneous substances and show that phase coexistence regions in high-dimensional Internal Energy space, $U(S,X_i,\ldots X_j)$, are simplicial convex polytopes—which are $N$-dimensional analogues of triangles and tetrahedra. In the first of this three-part series, we examine how the combinatorial relationships between the vertices and facets of simplicial polytopes leads to a generalized high-dimensional description of Gibbs' Phase Rule. Because Gibbs' Phase Rule describes the nature of phase boundaries on phase diagrams, this isomorphism between the physical principles of equilibrium thermodynamics and the geometry of simplicial polytopes provides the foundation to construct generalized phase diagrams, which can exist in any dimension, with any intensive or extensive thermodynamic variable on the axes.

### *New varieties of phase diagrams*

Between 1873 and 1876, Josiah Willard Gibbs wrote three seminal papers that established the geometric foundations of equilibrium phase diagrams.[1,2,3] Today, many tens of thousands of phase diagrams have been constructed and catalogued. However, despite the enormous number of phase diagrams now available, the variety of phase diagrams has remained fairly limited. Four main types of phase diagrams are in general use: 1) Temperature–Pressure; 2) Temperature–Composition; 3) Ellingham ($T$, $\mu_{O2}$); and 4) Pourbaix diagrams ($p$H, Redox potential). All four of these phase diagrams are derived from the Gibbs free energy, which has natural variables of temperature, pressure and composition. This thermodynamic potential is applicable to what Gibbs called 'simple systems'—defined to be macroscopically homogeneous, isotropic, uncharged and chemically inert; uninfluenced by gravity, electricity, distortion of the solid masses, or capillary tensions.

However, modern materials are decidedly *non-simple*. Today, we are becoming increasingly aware of how other forms of thermodynamic work—such as surface, [4,5,6] elastic, [7,8,9] electromagnetic, [10,11] electrochemical work,[73] *etc.*—influence phase stability under the complex chemical environments of modern materials devices. In **Table 1**, we list the conjugate intensive and extensive variables for various forms of thermodynamic work (also referred to as generalized forces and displacements, or field variables and molar quantities, respectively). **Figure 1** schematizes three representative cases where consideration of additional forms of thermodynamic work is essential:

**Table 1. Thermodynamic conjugate variables, as adapted from Reference 18**

| *Conjugate Variable* | *Intensive Variable* | *Extensive Variable* | *Differential term in dU* |
|---|---|---|---|
| **Thermal energy** | | | |
| Heat | Temperature ($T$) | Entropy ($S$) | $T\mathrm{d}S$ |
| **Mechanical work** | | | |
| Pressure-Volume | Pressure ($P$) | Volume ($V$) | $-\mathrm{PdV}$ |
| Stress-Strain | Stress ($\epsilon$) | Strain ($\sigma$) | $\sigma \mathrm{d}\epsilon$ |
| Gravitational | Gravitational Potential ($\psi = gh$) | Mass ($m = \Sigma M_i n_i$) | $\Psi \mathrm{d}m$ |
| Surface | Surface Energy ($\gamma$) | Surface Area ($A$) | $\gamma \mathrm{d}\underline{A}$ |
| **Electromagnetic work** | | | |
| Charge Transfer | Voltage ($\phi_i$) | Charge ($Q_i$) | $\phi_i \mathrm{d}Q_i$ |
| Electric Polarization | Electric Field ($\boldsymbol{E}$) | Dipole Moment ($\boldsymbol{P}$) | $\boldsymbol{E}\cdot d\boldsymbol{P}$ |
| Magnetic Polarization | Magnetic Field $\boldsymbol{B}$ | Magnetic Moment ($\boldsymbol{M}$) | $\boldsymbol{B}\cdot d\boldsymbol{M}$ |
| **Chemical work** | | | |
| Chemical reactions | Chemical Potential ($\mu_i$) | Component ($n_i$) | $\mu_i \mathrm{d}n_i$ |
| Solutions | Chemical Potential ($\mu_i$) | Mole fraction ($x_i$) | $\mu_i \mathrm{d}x_i$ |
| **Radiation Work** | | | |
| Electromagnetic | Wavelength ($hv$) | Photon Flux ($n$/Area) | $hv\cdot dn$ |
| Electrons/Particles | Kinetic Energy ($p^2$/2m) | Particle Flux ($n$/Area) | $p^2/2m\cdot dn$ |

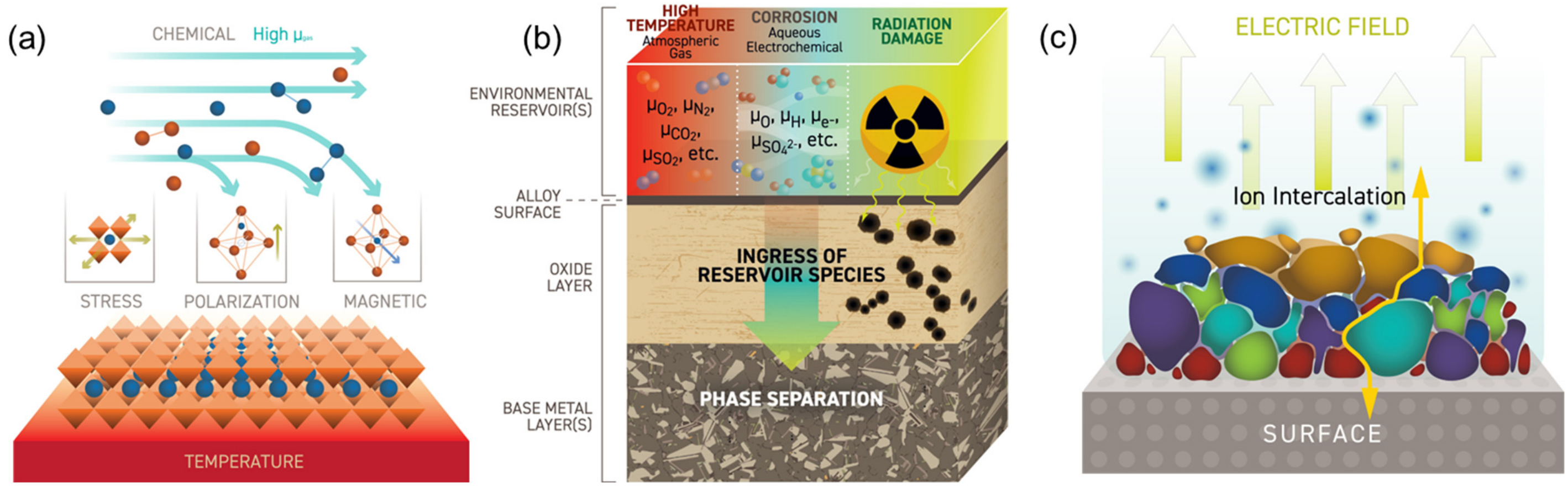

**Figure 1. Three representative examples where thermodynamic variables beyond temperature, pressure and composition are essential. a)** multiferroric materials, **b**) structural alloys in extreme chemical environments, **c**) electrochemical solid-liquid interfaces.

Multiferroric materials (**Figure 1a**) undergo symmetry-breaking phase transitions that couple ferroelectric ($E\cdot\boldsymbol{P}$) and/or ferromagnetic states ($B\cdot\boldsymbol{M}$) to elastic stresses and strains ($\sigma\epsilon$), and are a candidate class of materials for advanced sensors, actuators, and microelectronics.[10,11,12] Synthesizing multiferroric thin films with desired stoichiometry and phase relies on careful control of the fugacity ($\mu N$) and temperatures ($TS$) of the gaseous precursors. The polymorph that is deposited will be the phase with the lowest heterogeneous nucleation barrier, which is a function of the epitaxial strains ($\sigma\epsilon$) imposed onto the nucleus by the underlying substrate, as well as the interfacial energies ($\gamma A$) of the nucleating phase. Epitaxial strains ($\sigma\epsilon$) can also shift the morphotropic phase boundaries between polar and non-polar phases, or influence the switching barrier of ferroric phase transitions.

Structural metals (**Figure 1b**) usually contain numerous elemental constituents ($\mu_i x_i$)—either as dilute alloys in ferrous steels,[13] or as near-equimolar concentrations in so called 'high-entropy' alloys.[14,15] Designing the mechanical properties of alloys often relies on engineering their bulk (meta)stability,[16] such as by phase separation to form precipitates or interleaved eutectic microstructures. The stability of these alloys in extreme operational environments further depends on their reactions with a chemical reservoir—such as high-temperature air[17] or corrosive seawater[18]—which can ingress reservoir species ($\mu N$) and form interfacial oxide layers that degrade by exfoliation, spalling and comminution ($\gamma$A, $\sigma\epsilon$),[19] or form protective coatings with enhanced corrosion resistance ($\phi Q$) and mechanical wear resistance ($\sigma\epsilon$).[20] Radiation also appears to enhance phase stability in certain structural alloys,[21] or degrade stability by formation of voids and high internal surface area ($\gamma$A).[22]

The electrochemical solid-liquid interface (**Figure 1c**) governs the ion transport, interfacial stability and interphase formation between battery electrodes;[23] the nucleation and growth of minerals;[24,25] and the electrochemical stability and reaction kinetics of catalysts.[26] In ionic electrolytes, such as water and organic solvents, these solid-liquid interfaces can develop Electrical Double Layers[27] with complicated long-range heterogeneous atomistic structures, where the broken bonds of ionic solids ($\gamma$A) or electrically-charged metallic surfaces ($\Phi q$) drives chemisorption of charged ions ($\mu N$) that induces a persistent electric field ($E\cdot\boldsymbol{P}$), whose point of zero-charge can further be tuned by the $p$H ($N_{H+}\mu_{H+}$) and redox potential ($\phi Q$) of solvated ions and their concentrations ($\mu N$).[28,29]

These three examples represent an emerging crisis in modern materials science, where it is becoming increasingly important to build phase diagrams with axes beyond temperature, pressure and composition. Each additional form of work will increase the dimensionality of the free energy, by:

$$dU = TdS - PdV + \sum_i \mu_i dN_i + \gamma dA + \phi dQ + \sigma_{ijkl} d\varepsilon_{ijkl} + \vec{\boldsymbol{E}} \cdot d\vec{\boldsymbol{P}} + \vec{\boldsymbol{B}} \cdot d\vec{\boldsymbol{M}} + \ldots$$

This means that the relevant thermodynamic potential and phase diagram for these materials intrinsically reside in high-dimensional spaces. Because phase diagrams today are typically two dimensional, in general we have only considered lower-dimensional slices of these multi-dimensional phase diagrams, which results in a loss of essential thermodynamic information. This can lead to incorrect interpretations that certain observed metastable phases are 'non-equilibrium' states, when in fact these phases may be the lowest free-energy state somewhere on a higher-dimensional equilibrium phase diagram.[30,31,32,33] Even if non-equilibrium phases are indeed present, a high-dimensional description of their free energies will provide a more robust understanding of their kinetic origins.[34,35]

The time has come to lift ourselves out of the Flatland of 2D phase diagrams.[36] Urgently needed is a new theoretical framework that makes high-dimensional thermodynamics more accessible—both computationally and conceptually. The first sentence Gibbs wrote on thermodynamics reads:

> "*Although geometrical representations of propositions in the thermodynamics of fluids are in general use, and have done good service in disseminating clear notions in this science, yet they have by no means <u>received the extension in respect to variety and generality</u> of which they are capable*."

Written 150 years ago, Gibbs' call-to-action remains relevant today. To bring about the full capability of equilibrium thermodynamics, today we once again need to extend the variety and generality of thermodynamic phase diagrams.

In this series of three papers, we lay a geometric foundation to construct, interpret, and navigate generalized high-dimensional phase diagrams—which can be in any dimension, with any intensive or extensive thermodynamic variables on the axes. Although these three papers are certainly not the first to construct phase diagrams with various forms of work on the axes,[37,38,39,40,41,42] our emphasis here is to unify the geometric underpinnings of high-dimensional thermodynamics, which from our perspective, does not have a comprehensive and detailed framework in the literature yet.

On a high-dimensional phase diagram, stability regions and phase boundaries can also be high-dimensional objects. Because high-dimensional geometry is unfamiliar to our everyday experience, navigating high-dimensional phase diagrams can be difficult and unintuitive. This 3-part series is organized around a central thesis: that the conditions of heterogeneous equilibrium are represented geometrically by simplicial convex polytopes—which are *N*-dimensional analogues of triangles and tetrahedra.[43,44] The geometric properties of these simplicial polytopes—including their vertices, edges, and facets; as well as operations on convex polytopes—such as their slices, projections, duality, and gradients; offer a conceptual framework and computational toolkit to build and navigate high-dimensional phase diagrams. Our three papers are organized as follows:

1) **Phase diagrams with extensive variables as axes**: First, we revisit Gibbs' geometric arguments from his 1873 paper[3], "*A method of geometrical representation of the thermodynamic properties of substances by the means of surfaces.*" We extend the connection between Gibbs' arguments on heterogeneous equilibria with the modern mathematics of simplicial convex polytopes. We then show that the combinatorial geometry of vertices, edges and facets of simplicial convex polytopes give rise to rules and constraints governing phase coexistence regions on high-dimensional phase diagrams—from which we derive a generalized form of Gibbs' Phase Rule.

2) **Phase diagrams with intensive variables as axes:** Gibbs' arguments on the convex nature of equilibrium thermodynamics only applies to phase diagrams with extensive variables as axes. Next, we discuss how the Legendre transformation connects to the concept of Point-Line duality from projective geometry, which enables the computation of phase diagrams with all-intensive axes or mixed intensive/extensive axes. Using composition ($N$) and chemical potential ($\mu$) as example conjugate variables, we examine the dualities between closed and open thermodynamic boundary conditions, their corresponding thermodynamic potentials, and the resulting phase diagrams with axes of mixed intensive/extensive variables.

3) **Navigating high-dimensional phase diagrams**: Because high-dimensional phase diagrams are conceptually difficult to navigate, we next explore a concept of *relative stability*. If high-dimensional phase diagrams are a map, we need a 'compass' to point us in the direction that enhances (or reduces) the stability of specific phases. To formulate this compass, we derive a generalized high-dimensional Clausius-Clapeyron relation, which simultaneously identifies the full set of experimental parameters that stabilize or destabilize a particular phase of interest.

These three papers will necessarily suffer a similar limitation of Gibbs' original works, which is that the thermochemical data needed to showcase the full capability of high-dimensional phase diagrams is generally unavailable for any single chemical system. Like Gibbs, our discussions will focus primarily on the underlying thermodynamic formalism, rather than the thermochemical details of any specific chemical space. However, thermochemical data is becoming increasingly accessible due to recent advances in high-throughput density functional theory,[45] which has enabled the rapid and automated calculation of solid-state entropies,[46,47] equilibrium volumes,[48] strain tensors,[49] surface energies,[50,51] magnetic structures,[52] polarization displacements,[53] and other thermochemical properties from **Table 1**. By combining computed thermochemical data with our thermodynamic formalism, we can construct the high-dimensional phase diagrams that will revolutionize our design and understanding of materials in complex thermodynamic environments.

## Gibbs' derivation of heterogeneous equilibrium

Phase diagrams are scaffolded by thermodynamic axes, and mapped inside are stability regions for homogeneous pure phases, as well as the phase coexistence regions representing equilibrium heterogeneous mixtures. To generalize phase diagrams to higher dimensions, the key task is therefore to clarify the geometry of phase coexistence regions between two or more phases. This emphasis on determining the conditions of phase coexistence is why Gibbs titled his 1876 magnum opus: '*On the equilibrium of heterogeneous substances.*'[3]

Gibbs actually first described the geometry of phase coexistence in the second of his two 1873 papers,[2] titled, "*A method of geometrical representation of the thermodynamic properties of substances by the means of surfaces*." Here, we review the geometric arguments from this 1873 paper, using **Figure 2a** as an illustrative guide. First, we synthesize the essential physical arguments from both Gibbs' 1873 and 1876 papers, using modern thermodynamic notation,[54] to develop a generalized description of heterogeneous equilibrium that is agnostic to the form of thermodynamic work:

1. **The Energy Surface** Every substance has an Internal Energy, $U$, which can be represented as a function of extensive thermodynamic variables $X = S, V, N$, etc. The differential form of the Internal Energy is written as $dU = TdS + \Sigma Y_i dX_i$; where $Y_i = \partial U/\partial X_i$, and $Y_i X_i$ corresponds to the various thermodynamic conjugate variables listed in **Table 1**. The integrated form of this differential produces the Internal Energy surface, $U(S, V, N, etc)$. The Internal Energy surface is an intrinsically high-dimensional object, although in 'simple systems' only $S$, $V$, and $N$ terms are considered.

2. **Stability Criterion** For a homogeneous single-phase substance to be stable, its Internal Energy surface, $U$, must be positive-definite in all extensive variables; $\partial^2 U/\partial X^2 > 0$; or else the phase will self-separate by extent. Gibbs referred to the extensive state at which this convex curvature is violated as the *Limit of Essential Stability*, which we today refer to as the spinodal.

3. **Heterogeneous Equilibrium** A heterogeneous mixture of phases is at equilibrium when entropy is maximized, in other words, $\Delta S = 0$. Entropy is maximized when all intensive thermodynamic variables ($Y = T, P, \mu, etc.$) become equalized at the physical boundaries between phases within a heterogeneous mixture.

4. **Surfaces of Dissipated Energy** A system that has maximized entropy also has minimized Internal Energy. However, because each pure-phase $U(S,V,N,X...)$ surface is convex, at certain extensive conditions, a mixture of heterogeneous phases can have lower total Internal Energy than any single homogeneous phase. This state of heterogeneous equilibrium can be represented by the tangent plane that forms from the convex envelope connecting multiple $U(S,V,N\ldots)$ surfaces of pure phases.

   Gibbs referred to these tangent planes of phase coexistence as '*Surfaces of Dissipated Energy*.' The incline of these tangent planes along each extensive direction is $\partial U/\partial X_i$, which is by definition the conjugate intensive variable, $Y_i$. Because these tangent planes connect coexisting phases, this implies that coexisting phases have equalized intensive variables—which satisfies the condition of maximized entropy.

   Gibbs referred to the $U$ surface for pure homogeneous substances as the '*primitive surface*', and the convex envelope describing the state of heterogeneous equilibrium as the '*derived surface*'. In modern computational thermodynamics, the derived surface can be calculated with convex hull algorithms.[55,56] Phases that fall upon the derived surface are within the *limit of absolute* stability. As an example of a substance within the *limit of essential stability* but outside the *limit of absolute stability*, Gibbs described the metastable persistence of liquid water supercooled below 0°C.

5. **<u>The Phase Diagram</u>** The extensive states at which the *Surfaces of Dissipated Energy* connect with the *primitive surface* provide the boundaries between the single-phase and multi-phase regions of a phase diagram. In a multi-phase region, the phase fraction and extensive state of each phase can be solved by the lever rule. Projecting the *derived surface* onto axes of the thermodynamic variables produces the phase diagram. Lacking thermochemical data, Gibbs did not actually plot any phase diagrams himself, these were first constructed by J. D. van der Waals for binary solutions in 1890[57,58] and G. Tammann for water in 1900.[59]

Gibbs' nomenclature no longer appears in the modern thermodynamics curriculum, although concepts such as the *surface of dissipated energy* should remain familiar, particularly within the context of Gibbs free energy curves plotted against the extensive composition variable. However, it is not usually emphasized that the primitive and derived surfaces are, in fact, convex for *any* extensive variable from **Table 1**, due to Gibb's stability criterion (Argument 2, above). In fact, Gibbs' arguments regarding heterogeneous equilibrium were originally made in extensive Energy-Entropy-Volume ($U$-$S$-$V$) space, instead of the intensive $T$-$P$ axes shown on phase diagrams today. Gibbs was familiar with $P$-$V$-$T$ diagrams,[60] but deliberately made his arguments in extensive $S$ and $V$ space—explaining that if the relation between the volume, entropy and energy of body is known, then the relation between the energy, pressure, and temperature can be immediately deduced by differentiation; but the converse is not true, and thus knowledge of the former relation gives a more complete picture of the properties of a substance than knowledge of the latter.

James Clark Maxwell was so inspired by the graphical methods described in Gibbs' 1873 paper that he famously made a plaster model of $U(S,V)$ for a fictitious 'water-like' substance, which he mailed to Gibbs at Yale in 1875. We scanned and digitized this plaster model from the Yale Peabody Museum of Natural History,[61] and show in **Figure 2b** a colorized version of Maxwell's plaster model. **Figure 2b** can be interacted with in Augmented Reality using the QR code, and a 3D-printable model of Maxwell's surface is provided in the **Supplementary Files**. We also show the true $U(S,V)$ surface for water in **Figure 2c**, plotted using thermochemical data from the International Association for the Properties of Water and Steam.[62] (Note that in Figure 1c, volume is shown on a log-axis.) We also illustrate in **Figure 2d** the corresponding $S$-$V$ phase diagram for liquid water, ice-1h, and steam, formed by projecting the lower convex hull of the energy surface from **Figure 2c** onto the $S$ and $V$ axes.

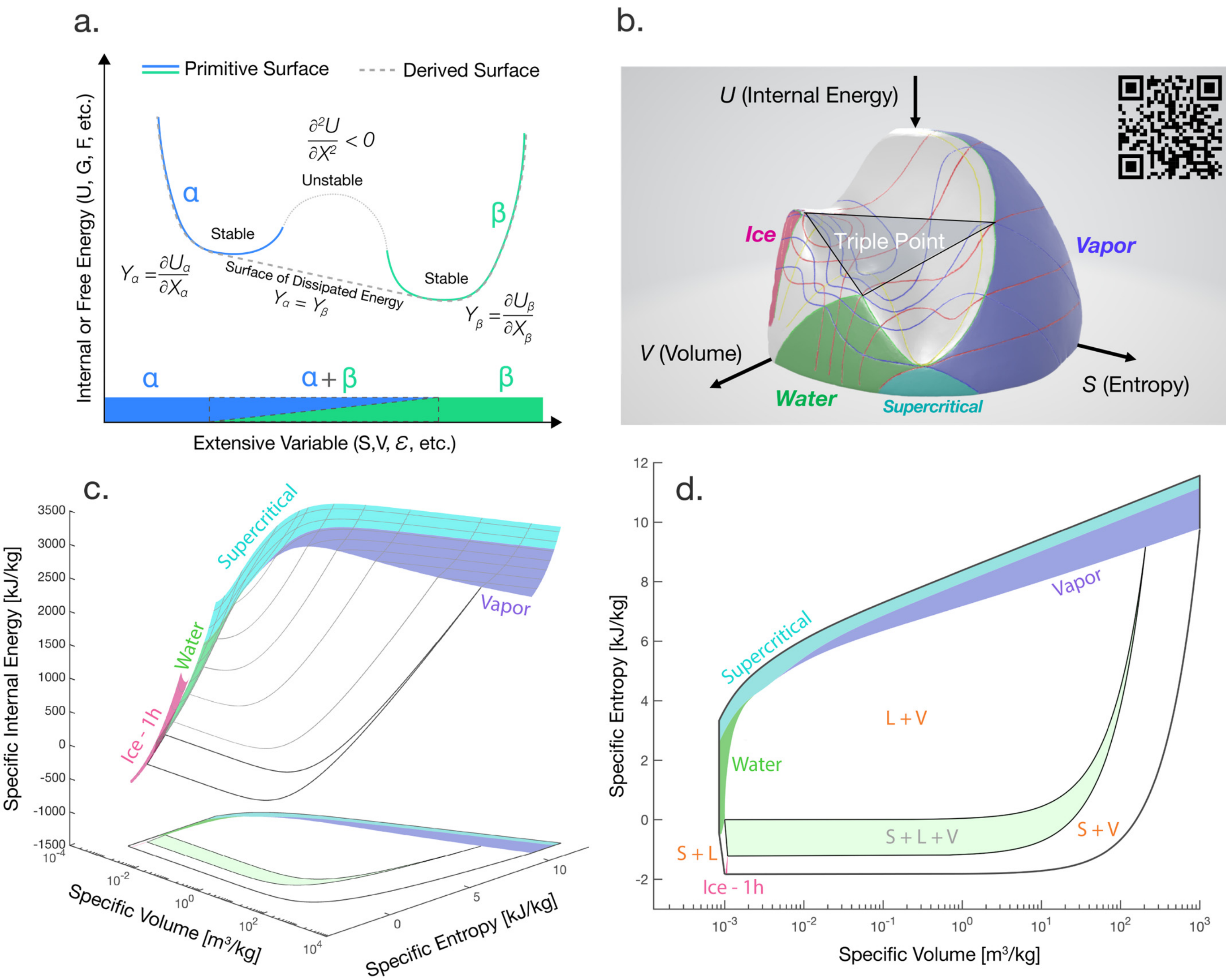

**Figure 2. Geometric origins of phase coexistence.** a) The *primitive* and the *derived* surfaces representing heterogeneous equilibrium. b) Colorized version of Maxwell's plaster model, showing the primitive $U(S,V)$ surface for 'water', as well as a triangle indicating the triple point for ice-water-vapor coexistence. The blue and red lines mark isobars and isotherms, respectively, and the yellow line marks the spinodal—referred to by Gibbs as the *Limits of Essential Stability*. Maxwell's surface can be interacted with in Augmented Reality by using a smartphone or tablet to view the QR code. c) The true $U(S,V)$ surfaces for water, ice-1h, and water vapor, calculated using the IAPWS-95 formulation. d) The $S$-$V$ diagram corresponding to the $U(S,V)$ energy surface shown in the previous panel.

On the Maxwell model, the equilibrium *derived surface* is constructed from the lower energy convex hull of the $U(S,V)$ surface. (Note that $U$ points down on Maxwell's model, such that up on the $z$-axis is negative in energy.) This *derived surface* for heterogeneous equilibrium can be mechanically simulated by rolling a tangent plane (such as a flat board) across the plaster model, where the gradient of the tangent plane, $\left\langle \frac{\partial U}{\partial S}, \frac{\partial U}{\partial V} \right\rangle$, gives the intensive conditions $\langle T, P \rangle$. When this tangent plane touches a pure single phase, the tangent plane can be tilted along both $\partial U/\partial S = T$ and $\partial U/\partial V = P$ axes without a phase transition, offering two tilt '*degrees-of-freedom*'. The tangent plane can also simultaneously touch two pure phases, but then one cannot vary $\partial$U/$\partial$S and $\partial$U/$\partial$V independently—there is now only one tilt degree-of-freedom. Finally, when the tangent plane simultaneously touches the 3 internal energy $U$(S,V) curves for ice, water and steam, this corresponds to a single $T$ and $P$ value for the coexistence of three phases—the triple point. There are zero tilt degrees-of-freedom at the triple point.

The number of single-phase $U$ curves that a tangent plane can touch simultaneously is the geometric origin of Gibbs Phase Rule. The relationship between the "degrees of freedom", $F$, and the number of coexistent phases, $P$, forms the language of Gibbs' Phase Rule, $F = C - P + 2$ (where for single-component water, $C = 1$). When the tangent plane touches multiple $U(S,V)$ curves simultaneously, the equilibrium state will be a heterogeneous mixture of multiple phases upon the *surface of dissipated energy*, with the phase fraction given by the lever rule.

## Phase Coexistence and Simplicial Convex Polytopes

In Gibbs' 1876 '*On the equilibrium of heterogeneous substances*', he briefly reviews the arguments from his 1873 paper in a section titled, "*Geometric Illustrations*". He then extends $U(S,V)$ to also include $N$—the extensive variable of chemical composition. Here, we extend Gibbs' arguments on heterogeneous equilibrium to arbitrarily many extensive variables, including strain, $\epsilon$; magnetic moment, $\boldsymbol{M}$; polarization, $\boldsymbol{P}$; electric charge, $Q$; and others from **Table 1**. Our generalization relies on the fact that $U(X)$ is convex with respect to <u>any</u> extensive variable—which highlights an essential similarity between extensive variables that arises from Gibbs' stability criterion. By accounting for these additional extensive variables, we pave the way to higher-dimensional phase diagrams, with thermodynamic variables beyond temperature, pressure and composition on the axes.

For each additional conjugate variable of thermodynamic work, the dimensionality of $U$ increases by one. It is difficult to visualize surfaces, convex hulls, and tangent hyperplanes in higher dimensions. Instead, to analyze the nature of phase coexistence under numerous operative forms of thermodynamic work, we must rely on dimensional analogy[36] and mathematical formalism. To do so, we now redevelop Gibbs' geometric arguments for heterogeneous equilibrium from the theory of high-dimensional convex polytopes. Not only does this geometric formalism help visualize phase coexistence in complex thermodynamic systems, we will show that theorems from the theory of convex polytopes can be used to derive new thermodynamic insights into the nature of phase coexistence.

We begin by defining that the thermodynamic surface $U$ exists in $\mathbb{R}^d$, which is a $d$-dimensional real Euclidean space. In $\mathbb{R}^d$, one of the dimensions is the scalar Internal Energy, and the relevant extensive variables comprise the other $(d–1)$ dimensions. The *primitive surface* is the $U$ surface of every pure homogeneous phase, and exists in $\mathbb{R}^d$ but is of dimensionality $(d–1)$. The *derived surface*, which corresponds to the equilibrium state and is formed from the lower convex hull of the primitive surface, also has a dimensionality of $(d–1)$.

Maxwell's $U(S,V)$ surface exists in $\mathbb{R}^3$, and the *surface of dissipated energy*—which connects the solid, liquid, and gaseous phases of water—produces a 2D triangle at the triple point. This triangle is also known as a 2-dimensional simplex, or '2-simplex'. In **Figure 2a**, where we were considering only one extensive variable, $U(X)$ exists in $\mathbb{R}^2$, so the surface of dissipated energy is a line, which is the 1-simplex. If another extensive variable, $X$, were included to form a $U(S,V,X)$ surface, the surface of dissipated energy in $\mathbb{R}^4$ would be a 3-simplex—a tetrahedron.

For an Internal Energy surface with $d-1$ conjugate thermodynamic variables that exists in $\mathbb{R}^d$, the *surface of dissipated energy* is a ($d$–1)-simplex. These phase-coexistence simplices touch the *primitive surface* of a homogeneous pure phase at a vertex. Therefore, **the maximum number of phases that can coexist in heterogeneous equilibrium is *d*,** which is the number of vertices on a ($d$–1)-simplex. In the language of convex polytopes, these high-dimensional phase-coexistence simplices are 'flat' hyperplanes (such as the flat 2D triangle in Maxwell's surface). In contrast, the *primitive surface* of homogeneous pure phases can be considered smooth, curved, and continuous manifold of 0-simplex vertices, such as for the pure single-phase regions in the Maxwell's surface.

A convex polytope where all the faces are flat simplices is known as a simplicial polytope. Simplicial polytopes are bounded by simplices of smaller dimension, known as $k$-faces, where $k$ is the dimension of the bounding simplex. For example, a 2-simplex (triangle) is bounded by three 1-simplices (edges) and three 0-simplices (vertices). The relationship between the numbers of $k$-faces of a simplicial polytope is given by the combinatorial geometry of the Dehn-Somerville equations, which is provided in **Table 2** for simplices up to 6 dimensions, and follows a pattern of Pascal's Triangle.

**Table 2 The combinatorial geometry of phase coexistence regions** is related to the number of $k$-faces of different dimension on a ($d$–1) simplex.

| | **Number of coexisting phases** | **1** | **2** | **3** | **4** | **5** | **6** | **7** |
|---|---|---|---|---|---|---|---|---|
| | Polytope face / $k$-simplex | 0-face (vertex) | 1-face (edges) | 2-face | 3-face | 4-face | 5-face | 6-face |
| **# of *N*-phase coexistence regions** | 0-*simplex* (point) | 1 | | | | | | |
| | 1-*simplex* (line segment) | 2 | 1 | | | | | |
| | 2-*simplex* (triangle) | 3 | 3 | 1 | | | | |
| | 3-*simplex* (tetrahedron) | 4 | 6 | 4 | 1 | | | |
| | 4-*simplex* (5-cell) | 5 | 10 | 10 | 5 | 1 | | |
| | 5-*simplex* | 6 | 15 | 20 | 15 | 6 | 1 | |
| | *6-simplex* | 7 | 21 | 35 | 35 | 21 | 7 | 1 |

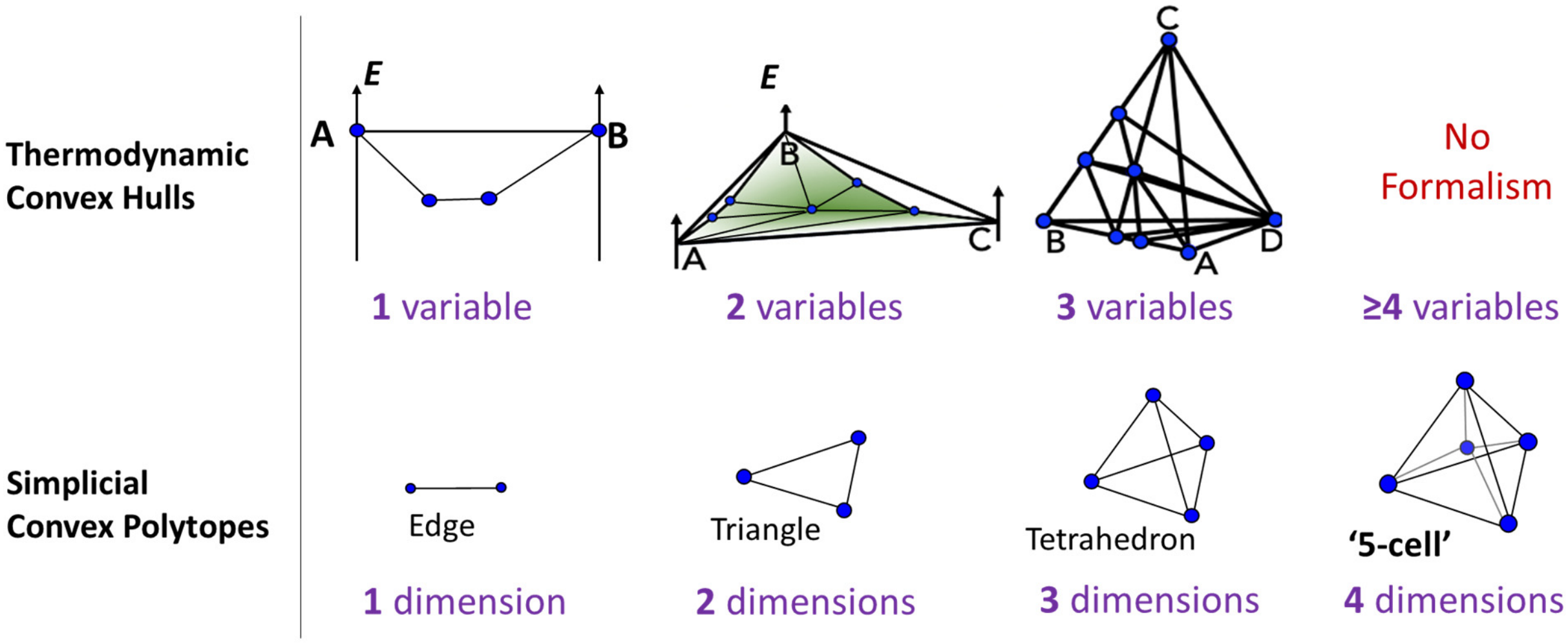

**Figure 3. Thermodynamic convex hulls have phase coexistence regions defined by flat simplicial polytopes**, with energy versus. composition convex hulls with 1, 2, 3, and 4 components (or other extensive variables).

Although the maximum number of phases that can coexist for ($d$ – 1) conjugate variables is $d$, the *derived surface* also exhibits phase boundaries along the $k$-faces as well. For example, on the Maxwell surface, the 2-simplex triangle which comprises the triple-point is bounded by two-phase coexistence *lines*, or 1-simplices, as visualized in **Figure 3**. These 1-simplices continuously connect the two-phase regions along the single tilt degree-of-freedom. This may be intuitive in 3-dimensional $U(S,V)$ space, but from **Table 2** we can also determine that a five-phase coexistence region on a four dimensional phase diagram is a '5-cell'—which is the 4-dimensional analogue of a triangle. In **Figure 3** we show a 5-cell projected into 2D. With some strenuous visualization, we can identify that a 5-cell is bounded by 5 tetrahedra (4-phase boundaries), 10 triangles (3-phase boundaries), 10 edges (2-phase boundaries) and 5 vertices (single homogeneous phases).

In this sense, all the phase-coexistence regions on an equilibrium derived surface are comprised of flat $k$-faces, with every dimensionality from $0 \leq k \leq d - 1$, where the $k = 0$ vertices are contact points with homogeneous pure-phase primitive surfaces, and $k > 0$ simplices represent heterogeneous equilibrium. For any $k$-simplex, the number of coexisting phases equals the number of vertices for that simplex, which is $k + 1$.

Not all convex polytopes are simplicial, for example, the faces on a cube have four vertices, so these square faces are not simplices. Although the only necessary condition for phase coexistence is that the *surfaces of dissipated energy* be convex polytopes, it is physically a near-certainty that these polytopes are also simplicial. For example, if the $U(S,V)$ surface on Maxwell's figure exhibited a fourth phase that was perfectly coincident with the same tangent plane as the triple-point triangle, then one could, in principle, have 4-phase coexistence in a single-component system. Gibbs wrote that this situation is 'entirely improbable' for real systems.[3] However, if one could tune interatomic interactions very precisely, one could, in principle, design a substance with this property. This was recently demonstrated computationally by Akahane *et al.* using a tunable Stillinger-Weber potential, achieving four-phase coexistence in a single-component material.[63] This phenomenon may perhaps also be achievable in the laboratory using a substance with precisely-tunable interaction parameters.

## Generalized Gibbs' Phase Rule

Simplicial convex polytopes offer a geometric description of phase boundaries in $U(S,X_i)$ space, which offers a pathway to generalize Gibb's Phase Rule to higher dimensions. As we visualized on Maxwell's thermodynamic surface, the '*degrees-of-freedom*' in Gibbs' Phase Rule refers to how many independent principal axes a tangent hyperplane can tilt along while remaining on the same phase-coexistence simplex. To fully define a flat ($d$–1) dimensional hyperplane, one needs $d$ linearly-independent vertices.[64] For example, to define a 2D hyperplane in $\mathbb{R}^3$, 3 vertices are required, which leaves zero tilt degrees of freedom. However, if this hyperplane in $\mathbb{R}^3$ is prescribed only by its tangency to a two-phase coexistence *line*, then the hyperplane retains 1 tilt degree-of-freedom about that line. Moreover, if the tangent hyperplane is prescribed by only its contact point with a pure-phase region, which corresponds to the 0-simplex in $\mathbb{R}^3$, then the hyperplane retains two tilt degrees of freedom.

Generalized Gibbs' Phase Rule can thus be derived by the following procedure: First, we determine the number of independent thermodynamic conjugate variable pairs to be considered, based either on interest or experimental context, which we call $W$. Per the preceding sections, the dimensionality of the Internal Energy surface in $\mathbb{R}^d$ is then $d = W + 1$. For $P$ coexisting phases, the dimensionality $k$ of their phase-coexistence simplex is $k = P - 1$. We next recall that the number of intensive degrees of freedom $F$ is equal to the tilt degrees of freedom of the tangent hyperplane. The phases (vertices) define the rank of the tangent hyperplane, and the degrees of freedom are the dimension of the null-space of the $k$-simplex connecting $P$ coexistent phases in $\mathbb{R}^d$, meaning $F = d - 1 - k$. **Generalized Gibbs' Phase Rule** is therefore: $\boldsymbol{F = W - P + 1}$, in which $F$ is number of degrees of freedom, $W$ is the number of independent thermodynamic conjugate variable pairs considered, and $P$ is the number of coexistent phases.

Although this result appears humble, its derivation is rooted in a careful consideration of the properties of high-dimensional complex polytopes. Its significance emerges in the fact that it can describe phase coexistence under applied conditions beyond temperature, pressure and composition. For example, consider the M1-Rutile polymorphism associated with the metal-insulator transition of $VO_2$. By Gibbs' Phase Rule, there should only be one stable polymorph at an arbitrary $T$ and $P$. However, when deposited on a substrate with enforced strains (an extensive variable), Generalized Gibbs Phase Rule says that $P = W + 1 - F$, where $\mathrm{d}U = T\mathrm{d}S + P\mathrm{d}V + \sigma\mathrm{d}\epsilon$ gives $W = 3$. $F = 2$ because $T$ and $P$ can be varied arbitrarily, and therefore $P = 2$. In other words, *two polymorphs can coexist in equilibrium under applied strains.* This two-phase coexistence was observed experimentally in the spontaneous decomposition of a strained $VO_2$ thin-film into stripes of M1/Rutile polymorphs.[65] Such observations of strain-based phase decomposition are conceptually identical to spinodal decomposition in the composition variables, but with an additional consideration of the tensorial nature of the stress-strain conjugate variables.[66,67]

Only *independent* thermodynamic conjugate variable pairs should be considered in evaluating $W$. In a single-component system for example, the chemical potential varies with the thermal ($TS$) and mechanical ($PV$) work variables by the Gibbs-Duhem equation,[68] and thus should not be incorporated into $W$. Likewise in a binary system, the chemical potentials of each component are not independent of each other (or of the thermal and mechanical work variables), and thus only a single compositional term is employed. We discuss this later in the "Components" section in more detail.

**Aside**: Throughout the history of thermodynamics, many have speculated on the relationship between the Euler Characteristic ($V - E + F = 2$) which describes the number of vertices, edges and faces on a 3D convex polytope, and Gibbs' Phase Rule ($F = C - P + 2$); as the two equations have similar form. The Dehn-Somerville equations provide a general description of the Euler Characteristic in $d$-dimensional simplicial polytopes, where the relationship between $k$-faces, $f_k$, is given as:

$$\sum_{j=k}^{d-1}(-1)^j\binom{j+1}{k+1}f_j=(-1)^{d-1}f_k$$

In three dimensions, the Dehn-Sommerville Equation becomes: $f_0 - f_1 + f_2 = 2$, where the subscript $k$ in $f_k$ indicates the dimensionality of the polytope facet (so $f_0$ = vertex, $f_1$ = edge, $f_2$ = triangular face, etc). In **Table 2**, the relationship between $k$-faces of higher dimension simplicial polytopes were shown. From our discussion, it is clear that the Euler characteristic refers to a different aspect of convex polytopes than Gibbs' Phase Rule, which is based on the tilt degrees of freedom of the tangent hyperplane. Although similar in form, and both deriving from the geometry of convex polytopes, the Euler Characteristic and Gibbs' Phase Rule are not directly related.

## Phase Diagrams with Intensive Natural Variables

The Internal Energy surface, $U$, is a fundamental and immutable property of substances, and does not vary under experimentally applied conditions. All the natural variables of the Internal Energy are extensive; however, it is conventional today to show phase diagrams with intensive variables, such as $T$ and $P$, which are often more convenient to experimentally measure and manipulate than their extensive conjugates. We now briefly discuss the relationship between these convex polytopes and phase diagrams of intensive natural variables, which we elaborate in much greater detail in Part II of this 3-part series.

To understand how a phase-coexistence simplex from $U$ appears on a phase diagram with intensive variable axes, we next examine how an envelope of planes tangent to $U$ interacts with these high-dimensional simplices. Consider again the 3-phase coexistence triangle from the Maxwell surface. If one rolls a tangent plane along the Maxwell surface with temperature fixed as a natural variable, then the plane can roll along only one principal axis of inclination, which is fixed to $T \equiv \left(\frac{\partial U}{\partial S}\right)_V$. If this $T = \partial U/\partial S|_V$ inclination of this rolling tangent plane is the same as the $\partial U/\partial S|_V$ inclination of the 3-phase coexistence triangle, in other words, if the temperature is chosen to be the triple-point temperature, then the triangle will appear collapsed from the vantage of the $T$-$V$ axes into a 1D line. The same 3 vertices from the 2D triangle will still be retained on this 1D line when collapsed.

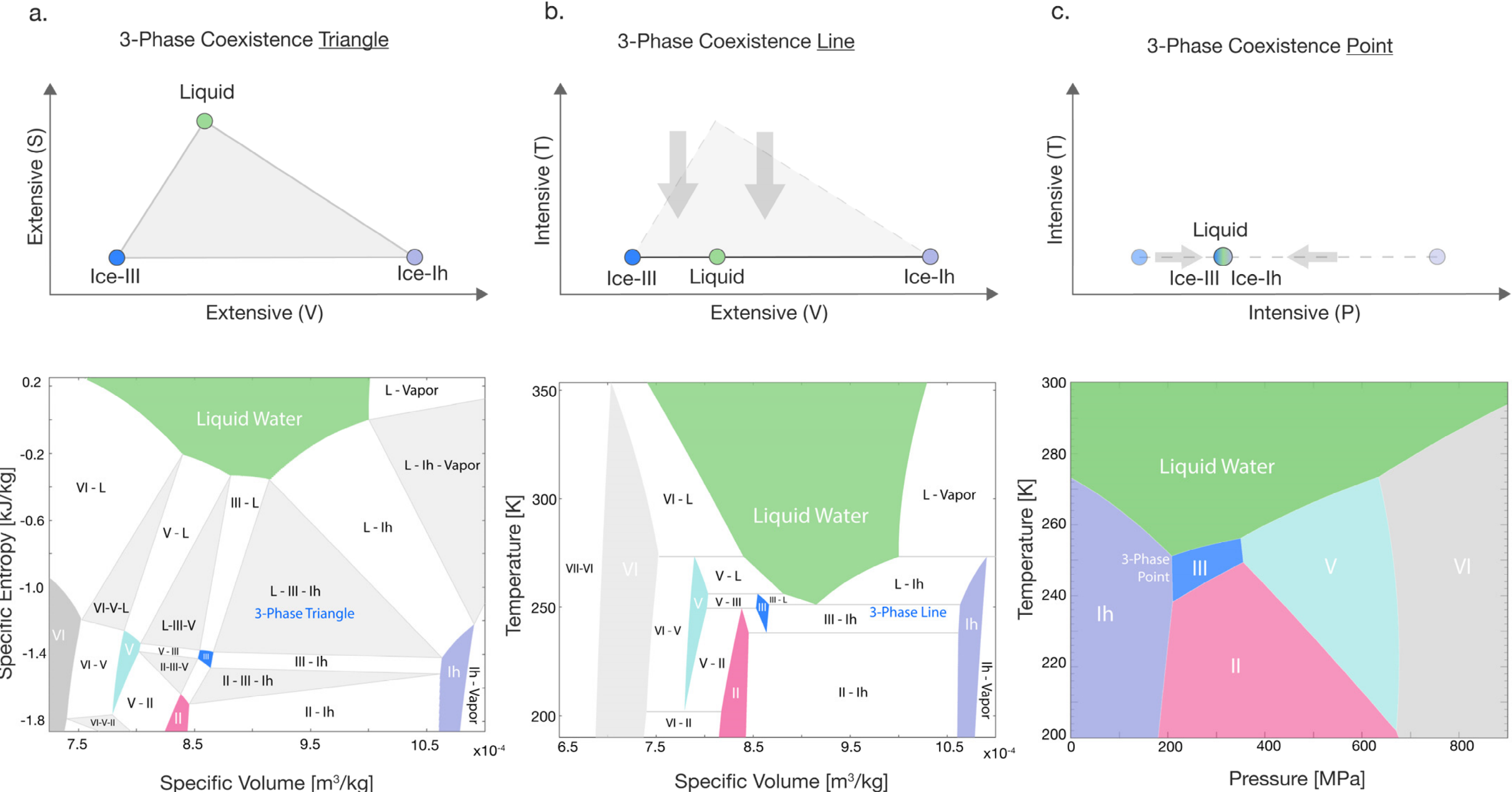

**Figure 4. Phase diagram simplex representations under a Legendre Transformation to an intensive natural variable.** Top) Simplex representations of three-phase coexistence and Bottom) phase diagrams of the condensed phases of $H_2O$ for **a.** two extensive natural variables (S,V) **b.** one extensive (V) and one intensive (T) natural variable **c.** two intensive natural variables (T,P).

**Figure 4** illustrates this concept of collapsed simplices, which shows the *S-V, T-V,* and *T-P* phase diagrams for the low-temperature, high-pressure polymorphs of $H_2O$. We examine this solid region of $H_2O$ so that the volume axis can be shown with linear scaling, as opposed to the logarithmic scaling in **Figure 1d**. In transforming from the *S-V* axes of **Figure 4a** to the *T-V* axes of **Figure 4b**, the three-phase coexistence triangles are collapsed in the *S* direction to produce three-phase invariant lines, which each retain the same three single-phase points that comprised the triangles. Shifting further to the *T-P* axes of **Figure 4c**, these three-phase coexistence lines are collapsed in the *V* direction to produce a three-phase coexistence point—the "triple point".

The *T-V* phase diagram of **Figure 4b** is isomorphic in many ways to traditional binary $G(T,x)$ phase diagrams, as analogous features to eutectoids and peritectoids appear, but in the extensive volume axis instead of the composition axis. This illustrates the fundamental geometric similarity between extensive variables in thermodynamics, which arises from the Gibbs' core arguments on the convexity of Internal Energy in the extensive variables.

Whether the natural variables are extensive or intensive *does not influence* phase coexistence from Gibbs' Phase Rule. Phase diagrams with intensive variables are simply dual representations of the Internal Energy surface produced by a Legendre transformation. However, based on the natural variables of interest and the resulting phase diagram, one might be inclined to distinguish between the 'extensive degrees of freedom', $F_X$, and regular intensive degrees of freedoms, $F_Y$. Here, an extensive degree of freedom refers to one's ability to change the value of an extensive variable and remain in the same phase coexistence region. In this case, the number of extensive degrees of freedom on a phase diagram are simply $F_X = W - I$, in which $I$ is the number of natural intensive variables. For example, in the $T$-$V$ diagram (for which where $W = 2$ and $I = 1$), the 3-phase invariance line exhibits 1 'extensive degree of freedom' along the volume axis.

## Components

Gibbs' Phase Rule places special emphasis on chemical components, which deals with the conjugate variables of $\Sigma\mu_i N_i$, where $N$ is the quantity of chemical component $i$. The nature of components has often produced confusion in the thermodynamics curriculum.[69] For example, is $CaCO_3$ a one, two, or three component material? Gibbs is not particularly helpful when defining what a component is, writing that the notion of a component may be determined "entirely by convenience, and independently of any theory in regard to the internal constitution of the mass." Composition is an extensive variable, so the number of components determines the dimensionality of the Internal Energy surface. Therefore, this confusion is important to resolve.

We can continue to interpret the nature of components from the geometry of convex polytopes. One important property of convex hulls is that ***any subset of a convex hull is also a convex hull***. When taking lower-dimensional subsections of a higher-dimensional convex hull, the stability of chemical compounds from the higher-dimensional convex hull is preserved. In **Figure 3a**, we show a hypothetical binary $N_A$-$N_B$ space, where $A$ and $B$ are elements. We plot the absolute Gibbs free energies of several hypothetical $A_xB_y$ compounds as red arrows, where $G = \mu_A N_A + \mu_B N_B$, which is extensive with respect to $N_A$ and $N_B$. States of equilibrium between these compounds are bounded by the lower convex hull around these phases (red arrows) in $G$-$N_A$-$N_B$ space.

For components, it is customary to represent compounds using an affine convex hull, where $x_i = N_i / N_{total}$ and $\Sigma x_i = 1$. In **Figure 5a**, this is illustrated by the purple slice taken along the direction $x_A = 1 - x_B$. This coupling between two extensive variables represents a change-of-basis from a $G$-$N_A$-$N_B$ space to a $G$-$x_A$-$N_{total}$ space. Chemical equilibrium does not depend on $N_{total}$, so components are usually analyzed in $G$-$x_A$ space, which eliminates the extensive variable $x_B$ and importantly, reduces the dimensionality of the thermodynamic space by one.

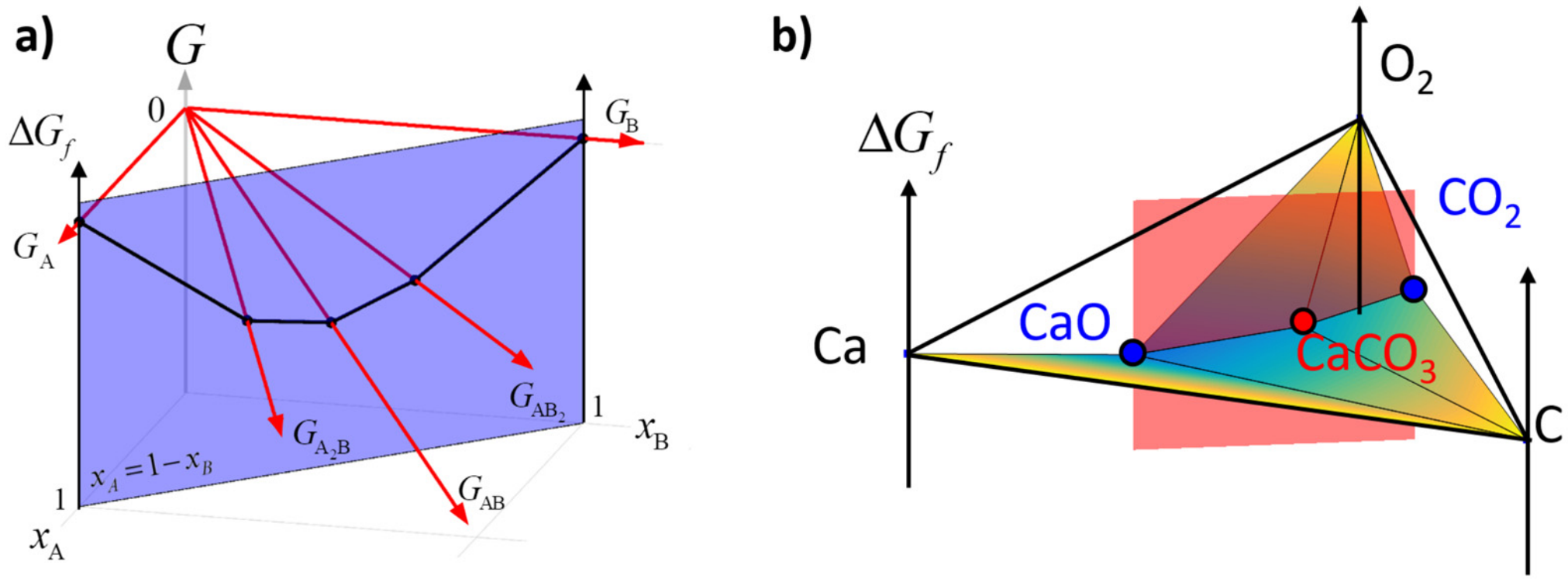

**Figure 5. a) A subset of a convex hull is also a convex hull.** The convex hull formed in G-$N_A$-$N_B$ space can be sliced along the $x_2 = 1 - x_1$ line, where $x_i = N_i/N_{total}$. The coupling of the extensive $x_1$ and $x_2$ variables reduces the dimensionality of the thermodynamic space by one. **b) The Ca-C-O elemental convex hull.** Compounds of Ca, C, and O can be analyzed as individual phases emerging from the ternary hull containing all three elements, as components around which to draw a lower-dimensional affine hull (as in the CaCO-$CO_2$, binary system marked by the pink plane), or as "pure" components in analysis of homogenous single-component systems ($CaCO_3$).

Chemical elements are the most elementary components, since they cannot be decomposed any further. As shown in **Figure 5b**, $CaCO_3$ is at most a system of three elemental components, which is represented by a 2-dimensional affine hull in composition space. However, by taking a subset of the 3-component convex hull along the CaO-$CO_2$ isopleth, $CaCO_3$ can be represented as a two-component material in this pseudo-binary convex hull. The single 0-simplex vertex of $CaCO_3$ is also a convex hull, and is a subset of the Ca-C-$O_2$ convex hull—meaning $CaCO_3$ can also be analyzed as a single-component material, for example, when studying its polymorphs. In principle, the 118 known elements form a 118-dimensional convex hull.[70] Any substance that can exist—inorganic solids, organic molecules, polymers, *etc.*—has a place within this 118-dimensional convex hull.

From the perspective of Generalized Gibbs Phase Rule, $\boldsymbol{F = W - P + 1}$, the minimum number of dimensions that the $\Sigma\mu_i N_i$ conjugate variables contributes to $W$ is provided by the smallest subset of this 118-dimensional convex hull that is relevant for a given experimental or theoretical analysis. This treatment is valid only within the domain of chemical potentials where the considered compounds do not decompose. For example, above 840°C, $\mu_{CO_2(g)} + \mu_{CaO} < \mu_{CaCO_3}$; and then it is no longer valid to treat $CaCO_3$ as a single-component material, it must be treated as at least a two-component material.

Aside from the coupling between chemical components to construct an affine hull, couplings between extensive quantities can also have other physical origins. For example, in the diffusion of a charged ion across a permeable membrane, as is often relevant in electrochemistry[71] and in biothermodynamics,[72] charge is coupled with mass (i.e. $dN$ is coupled with $zdN_{e^-}$, where $z$ is the charge of an ion). Similarly, in a Legendre transformation approach to the Pourbaix diagram, which is an electrochemical phase diagram for solid-aqueous systems,[73] the water reduction reaction also serves as an internal constraint, $H_2O \rightarrow \frac{1}{2} O_2 + 2H^+ + 2e^-$, such that $\mu_O = \mu_{H_2O} - 2 \cdot \mu_{H^+} + 2\Phi$.[74] Because coupled extensive variables can no longer be varied independently, the operative convex hull is now a reduced-dimensional subset of the original convex hull.

**New geometric constraints on phase equilibria**

Our generalization of Gibbs' Phase Rule as ***F = W – P + 1*** appears to be only a simple modification of the original Phase Rule. Similar modifications to the Phase Rule have been proposed before, but these previous derivations usually start with the Gibbs-Duhem relation, [75,76,77,78,79] whereas our discussion here was derived from the Maxwell surface and the geometry of high-dimensional complex polytopes.[2] When starting from the Gibbs-Duhem relation, it can be easy to overlook other interesting aspects of phase boundaries, which become apparent when the convex polytope geometry of phase boundaries is emphasized.

To illustrate this, we next leverage theorems from the mathematics of convex polytopes to reveal new insights regarding phase equilibria—specifically, we use the Upper Bound Theorem to establish fundamental constraints on the complexity of phase diagrams. Phase diagram complexity has important consequences in materials synthesis and design, as complex phase diagrams can lead to complicated materials synthesis pathways;[80,81,82] facile switching in multiferroric materials;[83,84] glass-formation in polymorphic systems,[85] and more. Understanding the geometric constraints on phase diagram complexity may therefore be a valuable consideration in materials design and development.

As discussed previously, the number of coexistent phases corresponds to the number of vertices of a convex simplicial polytope. For a given number of vertices, it is possible to produce simplicial polytopes with a varying numbers of *k*-faces,[86] or equivalently, varying numbers of phase-coexistence regions. However, the maximal number of *k*-faces that a simplicial polytope in $\mathbb{R}^d$ may contain for *n* vertices is bounded by McMullen's Upper Bound Theorem[87], and therefore, for a given number of phases, *P*, there are constraints on the maximal number of phase-coexistence regions on a phase diagram.

The Upper Bound Theorem asserts that the maximum number of *k*-faces of a convex polytope in $\mathbb{R}^d$ with *n* vertices is given by the number of *k*-faces on a cyclic polytope, $c(n,d)$; which is the polytope formed from the convex hull of *n* distinct points on a special curve in $\mathbb{R}^d$ named the rational normal curve. The equation for the Upper Bound Theorem is given as:

$$f_k(c(n,d)) = \sum_{r=0}^{\lfloor d/2 \rfloor} \binom{r}{d-k-1}\binom{n-d+r-1}{r} + \sum_{r=\lfloor d/2 \rfloor+1}^{\lfloor d \rfloor} \binom{r}{d-k-1}\binom{n-r-1}{d-r}$$

Where $d \geq 2$ and $0 \leq k \leq d-1$. (The floor notation $\lfloor x \rfloor$ means to round *x* down to the nearest integer.) Assuming that all phases have convex domains in extensive variable space, the Upper Bound Theorem thus establishes bounds on the maximum number of phase-coexistence regions in this space.

To demonstrate this principle, we analyzed the complexity of all ${}_{92}C_3$ = 125,580 ternary phase diagrams available on the Materials Project database.[88] **Figure 6a** shows the least complex and most complex phase diagrams as a function of the number of equilibrium phases, with complexity defined by the number of 3-phase coexistence regions. All listed phase diagrams can be explored on the free and publicly-available Materials Project website.

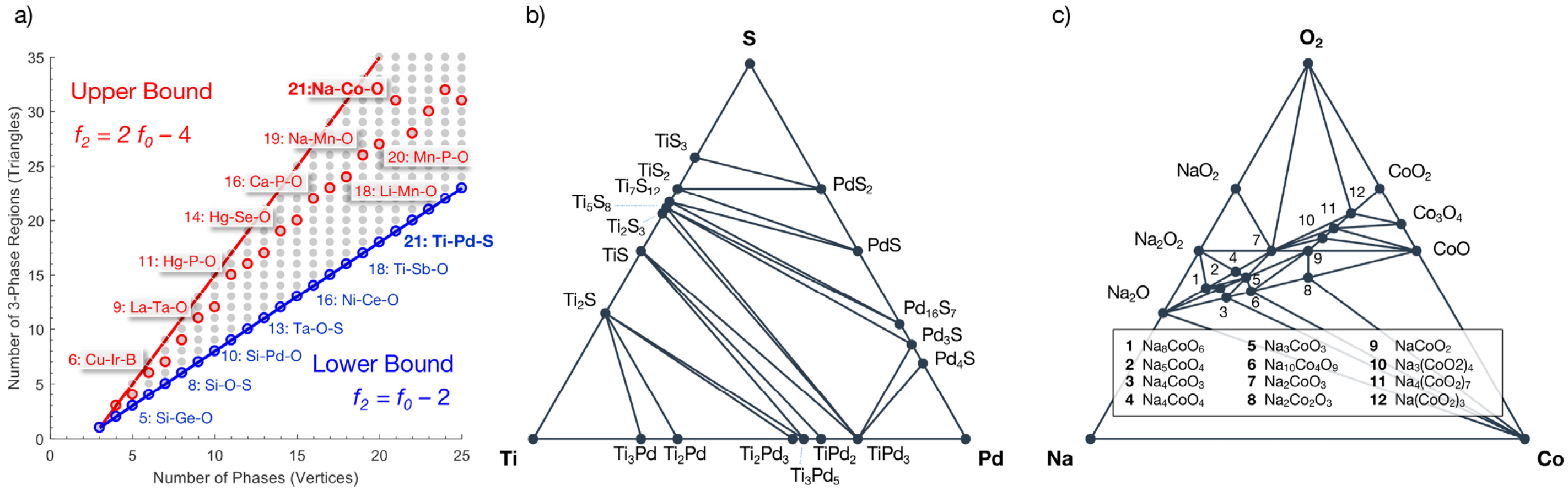

**Figure 6. Upper and lower bounds on ternary phase diagram complexity**. aa) Representative maximally and minimally complex phase diagrams for a given number of equilibrium phases is shown based on phase diagrams obtained from the Materials Project. For 21 equilibrium phases, b) Ti-Pd-S is the minimally complex ternary phase diagram whereas c) Na-Co-$O_2$ is the maximally complex.

The simplest ternary phase diagram is empty, formed by three elemental phases and 1 three-phase coexistence 2-simplex, with no equilibrium binary or ternary phases. For each additional phase, the minimal-complexity phase diagram results by splitting the triangle, adding one face for each additional equilibrium phase. Therefore, the lower bound complexity of a ternary phase diagram follows the relationship $f_2 = f_0 - 2$, where the notation $f_k$ indicates the number of faces of $k$-dimensionality. On the other hand, the Upper Bound Theorem states that the maximum complexity phase diagram follows a relationship $f_2 = 2 f_0 - 4$, and indeed there are no ternary phase diagrams that exceed this upper bound. For 21 equilibrium phases, we show in **Figure 6b** and **6c** the ternary phase diagram with minimum complexity, Pd-Ti-S; and maximum complexity, Na-Co-$O_2$.

On a broader note, by applying the Upper Bound Theorem to studying phase diagram complexity, we illustrate an exciting opportunity to examine how mathematical theorems from the rich literature of convex polytopes may translate into new physical insights regarding material thermodynamics. Conversely, problems in applied thermodynamics could stimulate new analyses within the mathematics of convex polytopes. This back-and-forth exchange will be fruitful to establish a new quantitative and conceptual foundation for analyzing the stability of modern materials.

**New Thermodynamic Variables**

Recently, it was demonstrated in a rod-polymer mixture that the rod aspect ratio and rod/polymer size ratio can be new extensive variables to describe 5-phase coexistence in a two-component colloidal system.[76] Radiation has also recently been shown to affect the nature of phase transformations, both light radiation as an electromagnetic wave,[89] as well as nuclear radiation in promoting the stability of high-entropy alloys.[90] Such observations raise a provocative question—is the list of thermodynamic conjugate variables in **Table 1** complete? Or are there other forms of thermodynamic work remaining to be discovered?

The final chapter of Callen's classic thermodynamic textbook[54] provides an illuminating discussion on the origins of thermodynamic variables. Equilibrium, as argued by Callen, is rooted in the time-translation symmetries of Noether's theorem. Phase transitions, then, arise from broken symmetries at a macroscopic level. The macroscopic features of matter that give rise to these broken symmetries can be applied as extensive thermodynamic variables. The criteria for these macroscopic features is qualified by Goldstone's Theorem: '*Any system with broken symmetry has a spectrum of excitations for which the frequency approaches zero as the wavelength becomes infinitely large'*. Callen describes how the extensive variable of volume originates from the zero frequency of a large-wavelength phonon mode; the electric/magnetic moment originates from the zero frequency of the large-wavelength dipole oscillation or magnetic spin waves; and the mole number arises from the gauge symmetry of fundamental particles. Plausibly, alignment of anisotropic colloidal crystals in a phase transition from isotropic or nematic phases to crystalline smectic phases could be interpreted as zero frequencies of the large-wavelength mode of uniaxial rod alignment, analogous to the alignment of electric or magnetic dipoles.

It seems entirely probable that there are other macroscopic features of matter that satisfy Goldstone's theorem still waiting to be identified, which could extend the list of thermodynamic conjugate variables from **Table 1**. New symmetry-breaking thermodynamic variables could fit into Landau's theory of critical phase transitions as order parameters[91], which account for configurational phase transitions analogous to the order-disorder of alloys, magnetic spins, and electric dipoles. Other thermodynamic variables may be simulated or contrived, such as the 'alchemical' parameters that vary interatomic interaction energy in computer simulations[92,93,94] or particle shape and symmetry.[95] Situations of broken time-symmetry, such as the recent experimental realization of Floquet phases of condensed matter;[96,97] or states of persistent heterogeneous equilibrium in dynamical but repeating non-equilibrium biological systems,[98,99] further question if *time* could become a phase diagram axis.

While one can, in principle, represent phase stability as a function of any variable, there are specific criteria that must be satisfied for a new variable, $X'$, to be an extensive thermodynamic variable that is consistent with Gibbs' convex hull-based principles of phase coexistence. These are: 1) The Internal Energy should scale extensively with $X'$ such that $U(\lambda X') = \lambda U(X')$; 2) there should be a conjugate variable $Y'=\partial U/\partial X'$ that is intensive, and the conjugate product $X'Y'$ should have units of energy; 3) A compound that is stable against self-separation should have an Internal Energy surface that is positive-definite with respect to $X'$, $\partial^2 U/\partial X'^2 > 0$; and 4) the resulting thermodynamic potential $dU = T\mathrm{d}S + Y'dX'$ should be an exact differential, representable in Pfaffian form.

The identification of truly new thermodynamic variables is an exciting prospect. Computationally, one could deploy convex hull algorithms to calculate new phase diagrams with these variables as axes, which could open up new avenues to control and manipulate matter. Taking thermodynamic partial derivatives[100] of the free-energy with respect to these new variables would produce new Maxwell relations; as well as new linear-response materials properties analogous to thermal expansion, magnetostriction, or bulk modulus. We are optimistic that provocative new thermodynamic variables await discovery, and that our mastery of them will advance the invention of new materials technologies.

## Conclusions

Physical theory advances when it is built upon a rigorous mathematical foundation. Here, we combined Gibbs' physical insights on the equilibrium of heterogeneous substances with the modern geometric formalism of convex polytopes, leading to a conceptual and computable framework to interpret and apply high-dimensional thermodynamics. Our derivation relies on an essential similarity in thermodynamic variables, where the Internal Energy, $U(S,X_i,\ldots)$, is convex with respect to any extensive variable. Consequently, the phase coexistence regions in high-dimensional Internal Energy space take the form of simplicial convex polytopes. This geometric description is a generalization of Maxwell's thermodynamic surface, but lifted up to higher-dimensional spaces where there are multiple axes of extensive thermodynamic work.

By identifying the relationship between simplicial convex polytopes and high-dimensional phase boundaries, a new opportunity emerges to examine how the combinatorial geometry, projections, slices, and dual representations of convex polytopes translate into new physical insights in equilibrium thermodynamics. Not only will this bring greater clarity and insight into the intrinsic high-dimensional structure of thermodynamics, it may give rise to new perspectives and strategies towards the design and synthesis of next-generation materials.

Finally, from this convex polytope description of phase boundaries, we derived a generalized form of Gibbs' Phase Rule, establishing the nature of phase boundaries when numerous forms of thermodynamic work are operative—such as surface, elastic, electrochemical, and electromagnetic work, and perhaps other forms of thermodynamic work that are not yet known today. In Part II of this three-part series, we will explore the point-line duality between $U(S,X)$ and its Legendre transformed thermodynamic potentials, as well as how to construct thermodynamic phase diagrams with mixed intensive/extensive axes, which correspond to thermodynamic systems with boundary conditions that are closed in some extensive quantities while being open in others. With Parts 1 and 2, we will have the geometric foundation to construct generalized high-dimensional phase diagrams, with any intensive or extensive variable on the axes.

## Supplementary Files

We provide Maxwell's thermodynamic surface as a Universal 3D Object (.u3d) file, which is the digitized and colorized version of the 3D-scanned surface from Yale Peabody Museum of Natural History.[61] The .u3d file can be viewed and edited in a variety of commercial and freeware 3D viewing and editing programs. The .u3d file can also be converted to an .obj or .stl file format, which can then be used for 3D printing.

## Acknowledgements

This work was supported as part of GENESIS: A Next Generation Synthesis Center, an Energy Frontier Research Center funded by the U.S. Department of Energy, Office of Science, Basic Energy Sciences under Award No. DESC0019212. WS thanks Dr. Alexi Baker from the Yale Peabody Museum of Natural History for producing the 3D digital scan of Maxwell's plaster model, and Talal Alothman from the Duderstadt Visualization Studio at the University of Michigan for preparing and colorizing Maxwell's surface colorizing for 3D printing and Augmented Reality.

## References

[1] Gibbs, J. W. Graphical Methods in the Thermodynamics of Fluids. *Transactions of the Connecticut* Academy, 2, (1873): 309-342

[2] Gibbs, Josiah Willard. "A method of geometrical representation of the thermodynamic properties by means of surfaces." *Transactions of Connecticut Academy of Arts and Sciences* (1873): 382-404.

[3] Gibbs, Josiah Willard. "On the equilibrium of heterogeneous substances." Transactions of the Connecticut Academy, III. pp. 108-248, (1876).

[4] Navrotsky, Alexandra. "Nanoscale effects on thermodynamics and phase equilibria in oxide systems." *ChemPhysChem* 12.12 (2011): 2207-2215.

[5] Frolov, Timofey, and Yuri Mishin. "Phases, phase equilibria, and phase rules in low-dimensional systems." *The Journal of chemical physics* 143.4 (2015).

[6] Elliott, Janet AW. "Gibbsian surface thermodynamics." *The Journal of Physical Chemistry B* 124.48 (2020): 10859-10878.

[7] Chen, Yimu, et al. "Strain engineering and epitaxial stabilization of halide perovskites." *Nature* 577.7789 (2020): 209-215.

[8] Schlom, Darrell G., et al. "A thin film approach to engineering functionality into oxides." *Journal of the American Ceramic Society* 91.8 (2008): 2429-2454.

[9] Fitzgerald, E. A., et al. "Influence of strain on semiconductor thin film epitaxy." *Journal of Vacuum Science & Technology A: Vacuum, Surfaces, and Films* 15.3 (1997): 1048-1056.

[10] Spaldin, Nicola A., and Manfred Fiebig. "The renaissance of magnetoelectric multiferroics." *Science* 309.5733 (2005): 391-392.

[11] Cheong, Sang-Wook, and Maxim Mostovoy. "Multiferroics: a magnetic twist for ferroelectricity." *Nature materials* 6.1 (2007): 13-20.

[12] Schlom, Darrell G., et al. "Strain tuning of ferroelectric thin films." *Annu. Rev. Mater. Res.* 37 (2007): 589-626.

[13] Olson, G. B. "Genomic materials design: The ferrous frontier." *Acta Materialia* 61.3 (2013): 771-781.

[14] George, Easo P., Dierk Raabe, and Robert O. Ritchie. "High-entropy alloys." *Nature reviews materials* 4.8 (2019): 515-534.

[15] Evans, Daniel, et al. "Visualizing temperature-dependent phase stability in high entropy alloys." *npj Computational Materials* 7.1 (2021): 151.

[16] Li, Zhiming, et al. "Metastable high-entropy dual-phase alloys overcome the strength–ductility trade-off." *Nature* 534.7606 (2016): 227-230.

[17] Robertson, J., and M. I. Manning. "Limits to adherence of oxide scales." *Materials Science and Technology* 6.1 (1990): 81-92.

[18] McCafferty, Edward. *Introduction to corrosion science*. Springer Science & Business Media, 2010.

[19] Robertson, J., and M. I. Manning. "Limits to adherence of oxide scales." *Materials Science and Technology* 6.1 (1990): 81-92.

[20] Borgioli, Francesca, et al. "Improvement of wear resistance of Ti–6Al–4V alloy by means of thermal oxidation." *Materials Letters* 59.17 (2005): 2159-2162.

[21] Kombaiah, Boopathy, et al. "Nanoprecipitates to enhance radiation tolerance in high-entropy alloys." *ACS Applied Materials & Interfaces* 15.3 (2023): 3912-3924.

[22] Kombaiah, Boopathy, et al. "Nanoprecipitates to enhance radiation tolerance in high-entropy alloys." *ACS Applied Materials & Interfaces* 15.3 (2023): 3912-3924.

[23] Wang, Luning, et al. "Identifying the components of the solid–electrolyte interphase in Li-ion batteries." *Nature chemistry* 11.9 (2019): 789-796.

[24] De Yoreo, James J., et al. "Crystallization by particle attachment in synthetic, biogenic, and geologic environments." *Science* 349.6247 (2015): aaa6760.

[25] De Yoreo, James J., Nico AJM Sommerdijk, and Patricia M. Dove. "Nucleation pathways in electrolyte solutions." *New perspectives on mineral nucleation and growth: from solution precursors to solid materials* (2017): 1-24.

[26] Gebbie, Matthew A., et al. "Linking Electric Double Layer Formation to Electrocatalytic Activity." *ACS Catalysis* 13.24 (2023): 16222-16239.

[27] Carnie, Steven L., and Glenn M. Torrie. "The statistical mechanics of the electrical double layer." *Advances in Chemical Physics* (1984): 141-253.

[28] Gulicovski, Jelena J., Ljiljana S. Čerović, and Slobodan K. Milonjić. "Point of zero charge and isoelectric point of alumina." *Materials and manufacturing processes* 23.6 (2008): 615-619.

[29] Ren, Yi, et al. "Effects of pH on Ice Nucleation by the α-Alumina (0001) Surface." *The Journal of Physical Chemistry C* 126.46 (2022): 19934-19946.

[30] Navrotsky, Alexandra. "Energetics at the nanoscale: Impacts for geochemistry, the environment, and materials." *MRS Bulletin* 41.2 (2016): 139-145.

[31] Sun, Wenhao, et al. "Thermodynamic routes to novel metastable nitrogen-rich nitrides." *Chemistry of Materials* 29.16 (2017): 6936-6946.

[32] Sun, Wenhao, et al. "Non-equilibrium crystallization pathways of manganese oxides in aqueous solution." *Nature communications* 10.1 (2019): 573.

[33] Sun, Wenhao, et al. "The thermodynamic scale of inorganic crystalline metastability." *Science advances* 2.11 (2016): e1600225.

[34] Bianchini, Matteo, et al. "The interplay between thermodynamics and kinetics in the solid-state synthesis of layered oxides." *Nature materials* 19.10 (2020): 1088-1095.

[35] Wang, Zheren, et al. "Optimal thermodynamic conditions to minimize kinetic by-products in aqueous materials synthesis." *Nature Synthesis* (2024): 1-10.

[36] Abbott, Edwin A. "Flatland: A romance of many dimensions." *Nineteenth Century Science Fiction*. Routledge, 1884. 57-68.

[37] Kaptay, George. "Nano-Calphad: extension of the Calphad method to systems with nano-phases and complexions." *Journal of materials science* 47 (2012): 8320-8335.

[38] Liu, Zi-Kui, and Yi Wang. *Computational thermodynamics of materials*. Cambridge University Press, 2016.

[39] Navrotsky, Alexandra. "Nanoscale effects on thermodynamics and phase equilibria in oxide systems." *ChemPhysChem* 12.12 (2011): 2207-2215.

[40] Xue, Fei, Yanzhou Ji, and Long-Qing Chen. "Theory of strain phase separation and strain spinodal: Applications to ferroelastic and ferroelectric systems." *Acta Materialia* 133 (2017): 147-159.

[41] Todd, Paul K., et al. "Selectivity in yttrium manganese oxide synthesis via local chemical potentials in hyperdimensional phase space." *Journal of the American Chemical Society* 143.37 (2021): 15185-15194.

[42] Voorhees, Peter W., and William C. Johnson. "The thermodynamics of elastically stressed crystals." *Solid state physics*. Vol. 59. Academic Press, 2004. 1-201.

[43] Grünbaum, Branko. *Convex polytopes*. Vol. 221. Springer Science & Business Media, 2013.

[44] Ziegler, Günter M. *Lectures on polytopes*. Vol. 152. Springer Science & Business Media, 2012.

[45] Curtarolo, Stefano, et al. "The high-throughput highway to computational materials design." *Nature materials* 12.3 (2013): 191-201.

[46] Petretto, Guido, et al. "High-throughput density-functional perturbation theory phonons for inorganic materials." *Scientific data* 5.1 (2018): 1-12.

[47] Wang, Yi, et al. "DFTTK: Density Functional Theory ToolKit for high-throughput lattice dynamics calculations." *Calphad* 75 (2021): 102355.

[48] Amsler, Maximilian, et al. "Exploring the high-pressure materials genome." *Physical Review X* 8.4 (2018): 041021.

[49] De Jong, Maarten, et al. "Charting the complete elastic properties of inorganic crystalline compounds." *Scientific data* 2.1 (2015): 1-13.

[50] Sun, Wenhao, and Gerbrand Ceder. "Efficient creation and convergence of surface slabs." *Surface Science* 617 (2013): 53-59.

[51] Tran, Richard, et al. "Surface energies of elemental crystals." *Scientific data* 3.1 (2016): 1-13.

[52] Horton, Matthew Kristofer, et al. "High-throughput prediction of the ground-state collinear magnetic order of inorganic materials using density functional theory." *npj Computational Materials* 5.1 (2019): 64.

[53] De Jong, Maarten, et al. "A database to enable discovery and design of piezoelectric materials." *Scientific data* 2.1 (2015): 1-13.

[54] Callen, Herbert B., and H. L. Scott. "Thermodynamics and an Introduction to Thermostatistics." (1998): 164-167.

[55] Ong, Shyue Ping, et al. "Li− Fe− P− O2 phase diagram from first principles calculations." *Chemistry of Materials* 20.5 (2008): 1798-1807.

[56] Liu, Zi-Kui. "Computational thermodynamics and its applications." *Acta Materialia* 200 (2020): 745-792.

[57] Waals, JD van der. "Molekulartheorie eines Körpers, der aus zwei verschiedenen Stoffen besteht." *Zeitschrift für Physikalische Chemie* 5.1 (1890): 133-173.

[58] Calado, Jorge CG, and JN Canongia Lopes. "The building-up of phase diagrams." *Pure and applied chemistry* 71.7 (1999): 1183-1196.

[59] Tammann, Gustav. "Ueber die Grenzen des festen Zustandes IV." *Annalen der Physik* 307.5 (1900): 1-31.

[60] Hastings, Charles Sheldon. *Biographical Memoir of Josiah Willard Gibbs, 1839-1903: By Charles S. Hastings*. Vol. 6. National Academy of Sciences, 1909.

[61] Object HST.290012 from the Division of the History of Science and Technology at Yale Peabody Museum of Natural History. 3D scanning by Chelsea Graham in the digitization lab of Yale's Institute for the Preservation of Cultural Heritage. https://collections.peabody.yale.edu/search/Record/YPM-HST-290012.

[62] International Association for the Properties of Water and Steam. Guideline on Thermodynamic Properties of Supercooled Water (IAPWS, 2015). http://www.iapws.org/relguide/IAPWS95-2018.pdf.

[63]Oses, Corey, Cormac Toher, and Stefano Curtarolo. "High-entropy ceramics." *Nature Reviews Materials* 5.4 (2020): 295-309.

[64] Akahane, Kenji, John Russo, and Hajime Tanaka. "A possible four-phase coexistence in a single-component system." *Nature communications* 7.1 (2016): 12599.

[65] Liu, M. K., et al. "Anisotropic electronic state via spontaneous phase separation in strained vanadium dioxide films." *Physical review letters* 111.9 (2013): 096602.

[66] Xue, Fei, et al. "Strain phase separation: Formation of ferroelastic domain structures." *Physical Review B* 94.22 (2016): 220101.

[67] Van de Walle, Axel. "A complete representation of structure–property relationships in crystals." *Nature materials* 7.6 (2008): 455-458.

[68] Sun, Wenhao, et al. "The thermodynamic scale of inorganic crystalline metastability." *Science advances* 2.11 (2016): e1600225.

[69] Alper, Joseph S. "The Gibbs phase rule revisited: Interrelationships between components and phases." *Journal of chemical education* 76.11 (1999): 1567.

[70] Hegde, Vinay I., et al. "The phase stability network of all inorganic materials." *Science advances* 6.9 (2020): eaay5606.

[71] Van der Ven, Anton, et al. "Rechargeable alkali-ion battery materials: theory and computation." *Chemical reviews* 120.14 (2020): 6977-7019.

[72] Alberty, R. A. "Thermodynamics of systems of biochemical reactions." *Journal of theoretical biology* 215.4 (2002): 491-501.

[73] Pourbaix, M. "Atlas of Chemical and Electrochemical Equilibria in Aqueous Solutions." *National Association of Corrosion Engineers: Houston, TX, USA* (1974).

[74] Sun, Wenhao, et al. "Non-equilibrium crystallization pathways of manganese oxides in aqueous solution." *Nature communications* 10.1 (2019): 573.

[75] Frolov, Timofey, and Yuri Mishin. "Phases, phase equilibria, and phase rules in low-dimensional systems." *The Journal of chemical physics* 143.4 (2015).

[76] Peters, V. F. D., et al. "76 phase rule: Evidence for an entropy-driven quintuple point in colloid-polymer mixtures." *Physical Review Letters* 125.12 (2020): 127803.

[77] Hillert, M. "Gibbs' phase rule applied to phase diagrams and transformations." *Journal of phase equilibria* 14 (1993): 418-424.

[78] Voorhees, Peter W., and William C. Johnson. "The thermodynamics of elastically stressed crystals." *Solid State Physics-Advances in Research and Applications*. Academic Press Inc, 2004. 1-201.

[79] Scott, Robert L. "Modification of the phase rule for optical enantiomers and other symmetric systems." *Journal of the Chemical Society, Faraday Transactions 2: Molecular and Chemical Physics* 73.3 (1977): 356-360.

[80] Alberty, Robert A. "Use of Legendre transforms in chemical thermodynamics (IUPAC Technical Report)." *Pure and Applied Chemistry* 73.8 (2001): 1349-1380.

[81] Bianchini, Matteo, et al. "The interplay between thermodynamics and kinetics in the solid-state synthesis of layered oxides." *Nature materials* 19.10 (2020): 1088-1095.

[82] Chen, Jiadong, et al. "Navigating phase diagram complexity to guide robotic inorganic materials synthesis." *arXiv preprint arXiv:2304.00743* (2023).

[83] Moya, X., Sohini Kar-Narayan, and Neil David Mathur. "Caloric materials near ferroic phase transitions." *Nature materials* 13.5 (2014): 439-450.

[84] Liu, Z. K., Xinyu Li, and Q. M. Zhang. "Maximizing the number of coexisting phases near invariant critical points for giant electrocaloric and electromechanical responses in ferroelectrics." *Applied Physics Letters* 101.8 (2012).

[85] Ferlat, Guillaume, et al. "Hidden polymorphs drive vitrification in B2O3." *Nature materials* 11.11 (2012): 925-929.

[86] Ziegler, Günter M. "Convex polytopes: extremal constructions and f-vector shapes." *arXiv preprint math/0411400* (2004).

[87] McMullen, Peter. "The maximum numbers of faces of a convex polytope." *Mathematika* 17.2 (1970): 179-184.

[88] Jain, Anubhav, et al. "Commentary: The Materials Project: A materials genome approach to accelerating materials innovation." *APL materials* 1.1 (2013).

[89] Oshima, Yu, Atsutomo Nakamura, and Katsuyuki Matsunaga. "Extraordinary plasticity of an inorganic semiconductor in darkness." *Science* 360.6390 (2018): 772-774.

[90] Zeng, Yinping, Yong Du, and Rainer Schmid-Fetzer. "Thermodynamics Under External Fields." *Journal of Phase Equilibria and Diffusion* 44.4 (2023): 539-541.

[91] Landau, Lev Davidovich, and Evgeniĭ Mikhaĭlovich Lifshitz. *Course of theoretical physics*. Elsevier, 2013.

[92] Damasceno, Pablo F., Michael Engel, and Sharon C. Glotzer. "Predictive self-assembly of polyhedra into complex structures." *Science* 337.6093 (2012): 453-457.

[93] Akahane, Kenji, John Russo, and Hajime Tanaka. "A possible four-phase coexistence in a single-component system." *Nature communications* 7.1 (2016): 12599.

[94] Sheppard, Daniel, Graeme Henkelman, and O. Anatole von Lilienfeld. "Alchemical derivatives of reaction energetics." *The Journal of chemical physics* 133.8 (2010).

[95] Frenkel, Daan. "Order through entropy." *Nature materials* 14.1 (2015): 9-12.

[96] Syrwid, Andrzej, Jakub Zakrzewski, and Krzysztof Sacha. "Time crystal behavior of excited eigenstates." *Physical Review Letters* 119.25 (2017): 250602.

[97] Zhang, Jiehang, et al. "Observation of a discrete time crystal." *Nature* 543.7644 (2017): 217-220.

[98] Takatori, Sho C., and John F. Brady. "Towards a thermodynamics of active matter." *Physical Review E* 91.3 (2015): 032117.

[99] Solon, Alexandre P., et al. "Generalized thermodynamics of phase equilibria in scalar active matter." *Physical Review E* 97.2 (2018): 020602.

[100] Gilmore, Robert. "Thermodynamic partial derivatives." *The Journal of Chemical Physics* 75.12 (1981): 5964-5966.